\documentclass[superscriptaddress,twocolumn,showpacs,prb]{revtex4-1}
\usepackage[utf8]{inputenc}
\usepackage{amsmath}
\usepackage{braket}
\usepackage{graphicx}
\usepackage{amsfonts}
\usepackage{pgfplots}
\usepackage{csquotes}
\usepackage{hhline}
\usepackage{amssymb}
\usepackage{listings}
\usepackage{color}
\usepackage{ulem}
\makeatletter
\newcommand{\ssymbol}[1]{^{\@fnsymbol{#1}}}
\makeatother

\usepackage{lipsum}
\definecolor{codegreen}{rgb}{0,0.6,0}
\definecolor{codegray}{rgb}{0.5,0.5,0.5}
\definecolor{codepurple}{rgb}{0.58,0,0.82}
\definecolor{backcolour}{rgb}{0.95,0.95,0.92}
 
\lstdefinestyle{mystyle}{
    backgroundcolor=\color{backcolour},   
    commentstyle=\color{codegreen},
    keywordstyle=\color{magenta},
    numberstyle=\tiny\color{codegray},
    stringstyle=\color{codepurple},
    basicstyle=\footnotesize,
    breakatwhitespace=false,         
    breaklines=true,                 
    captionpos=b,                    
    keepspaces=true,                 
    numbers=left,                    
    numbersep=5pt,                  
    showspaces=false,                
    showstringspaces=false,
    showtabs=false,                  
    tabsize=2
}
 
\usepackage{subfigure}

\usepackage{dcolumn}
\usepackage{tabularx}
\usepackage[colorlinks=true,linkcolor=blue,citecolor=blue,urlcolor=blue]{hyperref}
\usepackage{longtable}
\usepackage{braket}
\usepackage{float}
\newcolumntype{C}{>{\centering\arraybackslash}X}

\begin{document}
\title{Quantum networks using counterfactual quantum communication}

\author{Aakash Warke$^{a}$}
\email{warkeaakash@gmail.com}
\affiliation{Department of Physics, Johannes Gutenberg Universität Mainz, Institute of Physics, Staudingerweg 7, 55128 Mainz, Germany}

\author{Kishore Thapliyal}
\email{kishore.thapliyal@upol.cz}
\affiliation{Joint Laboratory of Optics, Faculty of Science, Palack\'{y} University, Czech
Republic, 17. listopadu 12, 771~46 Olomouc, Czech Republic}

\author{Anirban Pathak}
\email{anirban.pathak@gmail.com}
\affiliation{Department of Physics and Materials Science \& Engineering, Jaypee Institute of Information Technology, A 10, Sector 62, Noida UP-201309, India}
                                              
\begin{abstract}
Counterfactual quantum communication is one of the most interesting 
facets of quantum communication, allowing two parties to communicate without any transmission of quantum or classical particles between the parties involved in the communication process. This aspect of quantum communication originates from the interaction-free measurements where the chained quantum Zeno effect plays an important role. Here, we propose a new counterfactual quantum communication protocol for transmitting an entangled state from a pair of electrons to two independent photons. Interestingly, the protocol proposed here shows that the counterfactual method can be employed to transfer information from house qubits to flying qubits. Following this, we show that the protocol finds uses in building quantum repeaters leading to a counterfactual quantum network, enabling counterfactual communication over a linear quantum network. 
\end{abstract}


\maketitle
\section{Introduction\label{CQC_Sec1}}
One of the most intriguing aspects of quantum mechanics is the superposition in tensor product space leading to entanglement. This particular facet of quantum mechanics forms a basis for many applications in quantum information, especially quantum teleportation, quantum dense-coding, quantum key distribution, and quantum networks\cite{NCCUP2000, BBPRL1993,BW1993,NSRMP2011}. Conventionally, to perform quantum teleportation, an entangled pair of photons is distributed among two parties, say Alice and Bob. To transmit a quantum state, say $\Ket{\Psi}$, Alice makes a Bell measurement on his qubit and another unknown quantum state that he wants to transmit to Bob. Once the procedure of Bell measurement is complete, the prior photon pair gets disentangled and Bob's photon state is concluded by knowing Alice's measurement result. Alice uses a classical channel to deliver this measurement result acquired after the Bell measurement. This is the idea of quantum teleportation. Remote State Preparation (RSP) is another scheme of teleportation in which the sender is aware of the quantum state that is to be transferred. This scheme is known to be a more efficient scheme as in certain situations, it can be realized using a lesser number of classical bits in the transmission channel  \cite{PPRA2000}. Due to this fact, many schemes of remote state preparation and its modified versions have been proposed \cite{BDSSTWPRL2001, XSSJPB2007, MCLIJTP2018, QXJPLA2019, LWPLA2003, AHPRA2003, SPPLA2013, CNANS2014, SPPLA2013, CNANS2014, SSBPQIP2015}. A few years ago, an unconventional scheme for communication was proposed in which a sender transmits information to the other party without transmitting any classical or quantum particle in the transmission channel. The origination of this research was based on the principle of interaction-free measurements, proposed by Elitzur and Vaidman in 1993 \cite{ACLVQIP1993}. The basic idea of interaction-free measurements is that if there exists an absorption object in one of the arms of a Mach-Zehnder interferometer, the photon's interference observed by a detector is destroyed despite the photon not being absorbed by the absorptive object. Thus, using interaction-free measurements, one can determine the existence of an object without actually interacting with the object. However, in the original scheme for interaction-free measurement proposed by Elitzur and Vaidman \cite{ACLVQIP1993}, the maximum theoretical efficiency of detection of the object without interacting with it was 50\%. This work was followed by a set of theoretical and experimental papers that increased this efficiency of interaction-free measurements up to approximately 100\% owing to the discrete form of quantum Zeno effect \cite{PKPRL1995, BMJMP1977, K98, KWM+99}. Following this, generation of entanglement was demonstrated counterfactually by allowing the absorptive object to be in a quantum superposition state of presence and absence\cite{PGKPRL1999,HAPRA2003}. Subsequently, a novel idea called the "chained" quantum Zeno effect was proposed to achieve counterfactual quantum computation, thus leading to the counterfactual implementation of Grover's search algorithm with an efficiency of nearly 100\% \cite{OHN2006}. Following this, an in-depth analysis of the counterfactual quantum key distribution scheme was made \cite{ZQYPRA2010,ZQYPRA2012,YLPRL2012,JLZPRA2013,GBLPL2012}.  This type of unconventional scheme for transferring information with no quantum or classical particle present in the transmission channel present between the two parties is known as "counterfactual quantum communication" \cite{LVIJQI2016}. Counterfactual quantum communication protocols have already achieved considerable attention \cite{HSPRL2013,QGOSA2014,LZMPRA2015,HSFP2016}. In 2018, an analysis of all four Bell states was performed with the help of the counterfactual controlled-not gate\cite{FZSR2018}. Recently, researchers have revisited the area of counterfactual quantum communication and showed that a current of angular momentum $L_zmod2\hbar$ which carries one bit of information can travel through the transmission channel, without the transmission of the particle itself \cite{YAPRL2020}. This is highly counterintuitive, which is why the progress in the field of counterfactual quantum communication is increasing continuously and this counterintuitive nature has also motivated us to perform the present study. \\
Before we state the problem, we wish to address here, it would be apt to note that an important part of quantum communication is the transmission of an entangled state from one place to another. This has applications in quantum repeaters, quantum sensing, and a lot of other domains. Keeping this in mind, as an attempt to take the field of counterfactual quantum communication forward, here we propose a new protocol for counterfactually transmitting an entangled state from one place to another, i.e., to transmit an entangled state without the transmission of any quantum or classical particle in the transmission channel which connects the parties to share the entanglement. This scheme is unique and possesses certain benefits over the conventional methods of quantum teleportation or remote state preparation of entangled states, as here, the transmission channel does not require any quantum or classical particle to pass through it. The work is also inspired by the idea of combining two rapidly developing and interesting facets of quantum communication: (i) counterfactual quantum communication and  (ii) quantum networks. In what follows, we provide a protocol (along with necessary mathematical proof) of counterfactual entanglement transmission using a counterfactual CNOT gate and apply that protocol to generate a quantum network using linear optics.

The rest of the paper is organized as follows. An overview of the concept of a counterfactual CNOT gate which is an important concept used in our protocol is presented in Section \ref{CQC_Sec2}. Subsequently, Section \ref{CQC_Sec3} is dedicated to describing the proposed protocol for counterfactual quantum transmission of an entangled state. Section \ref{CQC_Sec4} involves a brief discussion on the application of the proposed protocol in the realization of the quantum repeaters, leading to a quantum repeater protocol. Finally, the paper is concluded in  Section \ref{CQC_Sec5}.

\section{Counterfactual CNOT gate \label{CQC_Sec2}}



The concept of a fully counterfactual CNOT gate depends on the functionality of Horizontal (Vertical) - Chained Quantum Zeno Effect or H(V)-CQZE setup as given in Table \ref{TableI}. These input-output relations are offered by the setup of CQZE and are deemed crucial in the construction of a counterfactual CNOT gate and are discussed in detail in Ref. \cite{HSFP2016,FZSR2018}. To visualize this input-output map, let us consider that there are two parties involved in the process, namely Alice and Bob. Bob possesses the quantum absorptive object (control bit) that is in a superposition state of $\Ket{pass}$ (representing transmission through Bob's object) and $\Ket{block}$ (representing absorption by Bob's object), and Alice possesses a photon which is in a superposition of $\Ket{H}$ and $\Ket{V}$. To describe the working of this gate, we may note that the initial state of the combined system as per the description above would be\\

\begin{table}
\centering
\begin{tabular}{|c|c|c|} 
\hline
Bob's control & Alice's target & Output\\
electron & photon & \\
\hline & &\\

$\Ket{pass}$ & $\Ket{I}_p$ & $\Ket{I}_p$\\

 & $\Ket{I^{\dagger}}_p$ & $-$\\
& & \\
\hline & & \\ 
$\Ket{block}$ & $\Ket{I}_p$ & $\Ket{I^\dagger}_p$\\

& $\Ket{I^\dagger}_p$ & $\Ket{I}_p$\\
& & \\
\hline
\end{tabular}
\caption{This table represents operations of H(V)-CQZE setup. Here, I,I$^\dagger $ $\in$ $\{H,V\}$\cite{FZSR2018}. Bob's control electron determines the output of the polarization of the photon that is sent as an input, i.e., Alice's target qubit. Thus, this setup is useful to construct a counterfactual CNOT gate.}
\label{TableI}
\end{table}

\begin{align*}
    \Ket{\psi} &= (\alpha\Ket{\text{pass}} + \beta\Ket{\text{block}}) \otimes (\lambda\Ket{H} + \mu\Ket{V}) \\
    &= \lambda\alpha\Ket{\text{pass}}\Ket{H} + \alpha\mu\Ket{\text{pass}}\Ket{V} \\
    &\quad + \beta\lambda\Ket{\text{block}}\Ket{H} + \beta\mu\Ket{\text{block}}\Ket{V}
\end{align*} 

After application of the counterfactual CNOT gate, state $\Ket{\psi}$ will be transformed to
\begin{eqnarray}
    \Ket{\psi\prime}=\lambda\alpha\Ket{pass}\Ket{H}+\alpha\mu\Ket{pass}\Ket{V}\nonumber\\
    +\beta\lambda\Ket{block}\Ket{V}+\beta\mu\Ket{block}\Ket{H}\nonumber\\
    =\lambda(\alpha\Ket{pass}\Ket{H}+\beta\Ket{block}\Ket{V})\nonumber\\+\mu(\alpha\Ket{pass}\Ket{V}+\beta\Ket{block}\Ket{H}.
\label{CQC_eqn5}
\end{eqnarray}

We may now change the representation to a convenient binary representation using the following rules:\\
\begin{eqnarray}
\Ket{pass}\xrightarrow[]{}\Ket{0}_e\nonumber\\
\Ket{block}\xrightarrow[]{}\Ket{1}_e\nonumber\\
\Ket{H}\xrightarrow[]{}\Ket{0}_p\nonumber\\
\Ket{V}\xrightarrow[]{}\Ket{1}_p,
\label{CQC_eqn6}
\end{eqnarray}
where, the subscripts $e, p$ denote whether the bit corresponds to Bob's absorbing material, the electron, or Alice's photon respectively. Following this notation, the final output entangled state of the counterfactual-CNOT gate can be represented as:
\begin{eqnarray}
\Ket{\psi\prime}=\lambda(\alpha\Ket{0}_e\Ket{0}_p+\beta\Ket{1}_e\Ket{1}_p)\nonumber\\
    +\mu(\alpha\Ket{0}_e\Ket{1}_p+\beta\Ket{1}_e\Ket{0}_p).
\label{CQC_eqn7}
\end{eqnarray}

As we can see from the above equation, the output of a counterfactual-CNOT gate as explained above is an entangled state with varying parameters of $\lambda$ and $\mu$ which can be fixed depending on the kind of polarization one prefers to choose for Alice and Bob. In our protocol, we consider horizontally polarized photons with Alice and Bob for ease of understanding of the proof that counterfactual transmission of an entangled state is possible. 

\section{Protocol for counterfactual quantum transmission of an entangled state}\label{CQC_Sec3}
Now that a counterfactual CNOT gate has been described briefly, we may move forward to describe our protocol for transmitting an entangled state from one place to another without any transmission of quantum or classical particles in the transmission channel. We aim to transmit an entangled state counterfactually, for which an entangled pair of electrons is required. We approach this problem as follows. To begin with, we consider that Alice, Bob, and Charlie are the three parties involved in the protocol. Charlie possesses an entangled pair of electrons, whereas, at the same time, Alice and Bob each possess a horizontally polarized photon $\ket{H}$. Charlie uses two counterfactual-CNOT gate to entangle his pair of electrons to a pair of independent $\ket{H}$ photons possessed by Alice and Bob in such a way that one electron of Charlie gets entangled with the photon of Alice via the use of a  counterfactual-CNOT gate and the other one gets entangled with Bob's photon via the use of another counterfactual-CNOT gate. Since the counterfactual-CNOT gate is completely counterfactual, the entanglement transfer remains counterfactual. Now, to make sure that this entangled state is transmitted to the two photons, a Bell measurement is made on Charlie's electrons, which ultimately leads to the entanglement between Alice's and Bob's photons. Note that during this process, only Charlie operates on his pair of electrons while Alice's and Bob's photons are not interacting. Thus, the information in an entangled state which was with Charlie is now transmitted to two independent photons possessed by Alice and Bob using a counterfactual CNOT gate. In what follows, we provide a step-by-step mathematical description of the aforementioned idea as a protocol.

\begin{figure*}
    \centering
    \includegraphics[scale=0.36]{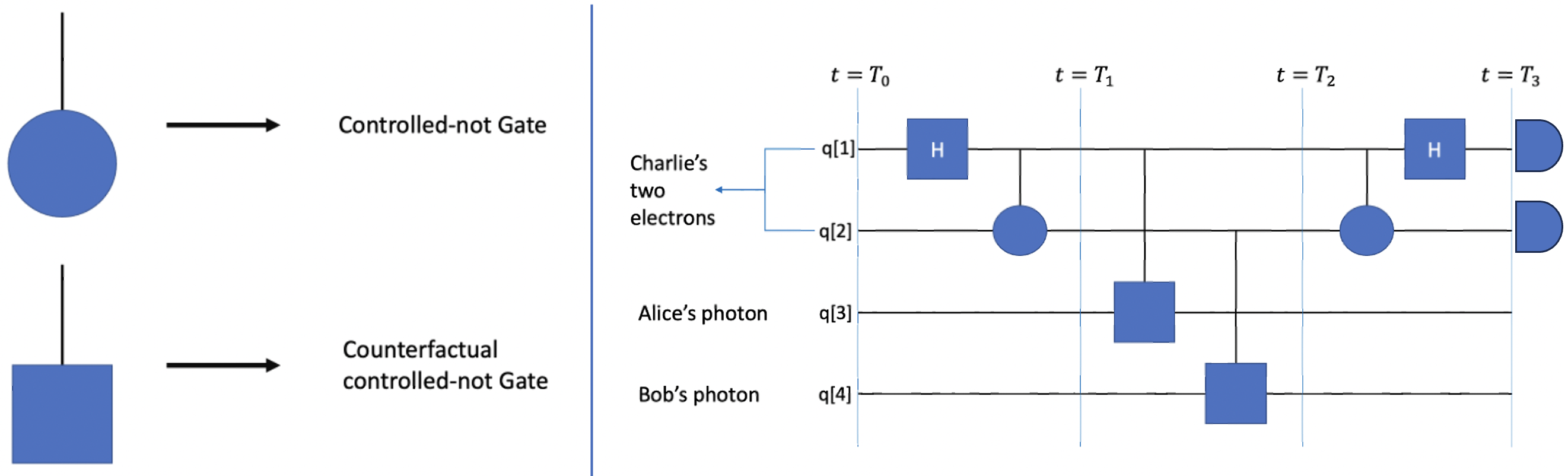}
    \caption{This figure represents our proposed circuit which inculcates the idea of counterfactually transmitting the information embedded in an entangled state of electrons. Counterfactual-CNOT, CNOT, and Hadamard are the three quantum gates involved in this circuit. The three parties involved in this protocol are Alice, Bob, and Charlie. Charlie possesses the first two qubits, q[1] and q[2] which are electrons. Alice and Bob possess one horizontally polarized photon each, at q[3] and q[4] respectively. At $t = T_0$, Charlie prepares an entangled state with his two electrons. We have considered it to be a maximally entangled state. However, in principle, any entangled state can be used for this purpose. At $t=T_1$, counterfactual-CNOT gates are used from q[1] to q[3] and q[2] to q[4]. This means that now a four-partite state is formed with Charlie's two electrons and Alice and Bob's horizontally polarized photons. At $t=T_2$, Bell measurement is made on Charlie's two electrons which ensures that Alice's and Bob's photons are entangled. In the entire process, two photons neither interact with each other nor there happens to be a transmission of a particle between the channel that connects Charlie to Alice and Bob or Alice to Bob.}
    \label{CQC_Fig1.png}
\end{figure*}

\begin{itemize}
    \item Referring to Fig. \ref{CQC_Fig1.png}, let's say at $t = T_0$, Charlie possesses qubit-1 and qubit-2, while Alice and Bob possess qubit-3 and qubit-4 respectively. Thus, the joint state of the system at $t = T_0$ is:
    \begin{eqnarray}
        \Ket{\Psi_0}=\Ket{g}_{e1}\Ket{g}_{e2}\ket{H}_{p1}\Ket{H}_{p2},
    \label{CQC_eqn8}
    \end{eqnarray}
    where, the subscripts $e1$, $e2$, $p1$ and $p2$ refer to the first electron with Charlie, the second electron with Charlie, the photon with Alice, and the photon with Bob respectively. $\Ket{g}$ refers to the ground state of the electrons and $\Ket{H}$ refers to the polarization state of the photon.
    
    \item At $t = T_1$, Charlie's electrons are entangled using a Hadamard and a CNOT gate. The information in this entangled state is what we aim to transmit counterfactually to a pair of independent photons without any prior contact between these two parties. Thus, at $t = T_1$, the joint state of the system would be:
    \begin{eqnarray}
        \Ket{\Psi_1}=\frac{1}{\sqrt{2}}\big(\Ket{g}_{e1}\Ket{g}_{e2}+\Ket{e'}_{e1}\Ket{e'}_{e2})\nonumber\\
        \otimes (\ket{H}_{p1}\Ket{H}_{p2}\big).
    \label{CQC_eqn9}
    \end{eqnarray}
    Here, $\Ket{e'}$ represents the excited state of the electron. Representing $\Ket{g}$ as $\Ket{block}$ and $\Ket{e'}$ as $\Ket{pass}$, we can represent the state as
    \begin{eqnarray}
        \Ket{\Psi_1}=\frac{1}{\sqrt{2}}\big(\Ket{pass}_{e1}\ket{H}_{p1}\Ket{pass}_{e2}\Ket{H}_{p2}\nonumber\\
        +\Ket{block}_{e1}\ket{H}_{p1}\Ket{block}_{e2}\Ket{H}_{p2}\big).
    \label{CQC_eqn10}
    \end{eqnarray}
    
    \item Since we have Alice's and Bob's photon's polarization states as $\Ket{H}$ only and not a superposition, it means that the value of $\lambda$ is 1 and the value of $\mu$ is 0 as per the explanation provided in Section \ref{CQC_Sec2}. Since the value of $\mu$ is equal to zero, we need not worry about the second term in Eq. \eqref{CQC_eqn5} and \eqref{CQC_eqn7}. Similarly, due to the symmetry of a Hadamard gate, $\alpha=\beta=\frac{1}{\sqrt{2}}$. At $t=T_2$, a counterfactual-CNOT gate is applied from $e1$ to $p1$, and another counterfactual-CNOT gate is applied from $e2$ to $p2$ which leads to the effective joint state as given below:
    \begin{eqnarray}
        \Ket{\Psi_2}=\frac{1}{\sqrt{2}}\big(\Ket{pass}_{e1}\ket{H}_{p1}\Ket{pass}_{e2}\Ket{H}_{p2}\nonumber\\
        +\Ket{block}_{e1}\ket{V}_{p1}\Ket{block}_{e2}\Ket{V}_{p2}\big).
    \label{CQC_eqn11}
    \end{eqnarray}
    Following the binary representations as shown in Eq. \eqref{CQC_eqn6}, we can represent Eq. \eqref{CQC_eqn11} as:
    \begin{eqnarray}
        \Ket{\Psi_2}=\frac{1}{\sqrt{2}}\big(\Ket{0}_{e1}\ket{0}_{p1}\Ket{0}_{e2}\Ket{0}_{p2}\nonumber\\
        +\Ket{1}_{e1}\ket{1}_{p1}\Ket{1}_{e2}\Ket{1}_{p2}\big).
    \label{CQC_eqn12}
    \end{eqnarray}
    
    Using the property of tensor products, we can again simply re-write Eq. \eqref{CQC_eqn12} as:
     \begin{eqnarray}
        \Ket{\Psi_2}=\frac{1}{\sqrt{2}}\big(\Ket{00}_{e1,e2}\Ket{00}_{p1,p2}
        +\nonumber\\\Ket{11}_{e1,e2}\Ket{11}_{p1,p2}\big).
    \label{CQC_eqn13}
    \end{eqnarray}
    
    \item To retrieve the information, Charlie needs to perform a Bell measurement on his two electrons $e1$ and $e2$. Eq. \eqref{CQC_eqn13} in terms of Bell states can be represented as:
    \begin{eqnarray}
        \Ket{\Psi_2}=\frac{1}{\sqrt{2}}\big(\Ket{\psi^+}_{e1,e2}\Ket{\psi^+}_{p1,p2}
        +\nonumber\\\Ket{\psi^-}_{e1,e2}\Ket{\psi^-}_{p1,p2}\big)
    \end{eqnarray}
    Here, $\Ket{\psi^\pm} = \frac{1}{\sqrt{2}}(\Ket{00}_i\pm\Ket{11}_i)$ where $i \in \{(e1,e2) , (p1,p2)\}$. At $t = T_3$, Charlie has performed a Bell measurement on his two electrons $e1$ and $e2$. Depending on the state that he acquires after his measurement, he would be able to conclude the kind of entanglement that is shared between Alice's and Bob's photons. The entanglement shared between Alice and Bob is going to be one among $\Ket{\psi^\pm}$ Bell states.\\
\end{itemize}
Thus, we have successfully achieved counterfactual quantum communication of an entangled state from a pair of electrons to a pair of independent photons possessed by two different parties. Such a scheme can be useful in developing counterfactual quantum key distribution protocols or possess applications in increasing the effectiveness of quantum communication through quantum repeaters. We discuss this application of quantum repeaters in the next section.\\

\section{Quantum Repeater Protocol \label{CQC_Sec4}}

During this process of entanglement transfer, there happens to be no trace of a quantum or a classical particle in the transmission channel that connects Charlie to Alice/Bob, as the counterfactual-CNOT gate is completely counterfactual. One can start with vertically polarized photons from Alice's and Bob's ends to show that the output entangled state will be cross-polarized as $\lambda$ would be equal to 0 whereas $\mu$ would be equal to 1 in this case. One can also take a superposition of horizontally and vertically polarized photons to obtain a more generic output. One can further prove that transmission of three-qubit GHZ states is also possible counterfactually by appropriate use of counterfactual-CNOT gates. Such schemes can be useful for long-distance quantum communication through fibers. As we know the field of quantum communication has gradually moved from experiments in laboratories to commercialization by industries \cite{QRJPC2019}, overtime, it is going to get highly important to realize the need for it on a global scale. Keeping this in mind, we put forward an application in the context of quantum repeaters using the aforementioned scheme. Quantum repeaters would be extremely useful in increasing the distance of communication while lowering the transmission loss in optical channels. The core idea of this is entanglement swapping, the possibility of realizing the same in a counterfactual manner has been shown in the previous section. 

\subsection{Working of the proposed quantum repeater protocol}
\begin{figure*}
    \centering
    \includegraphics[scale=0.43]{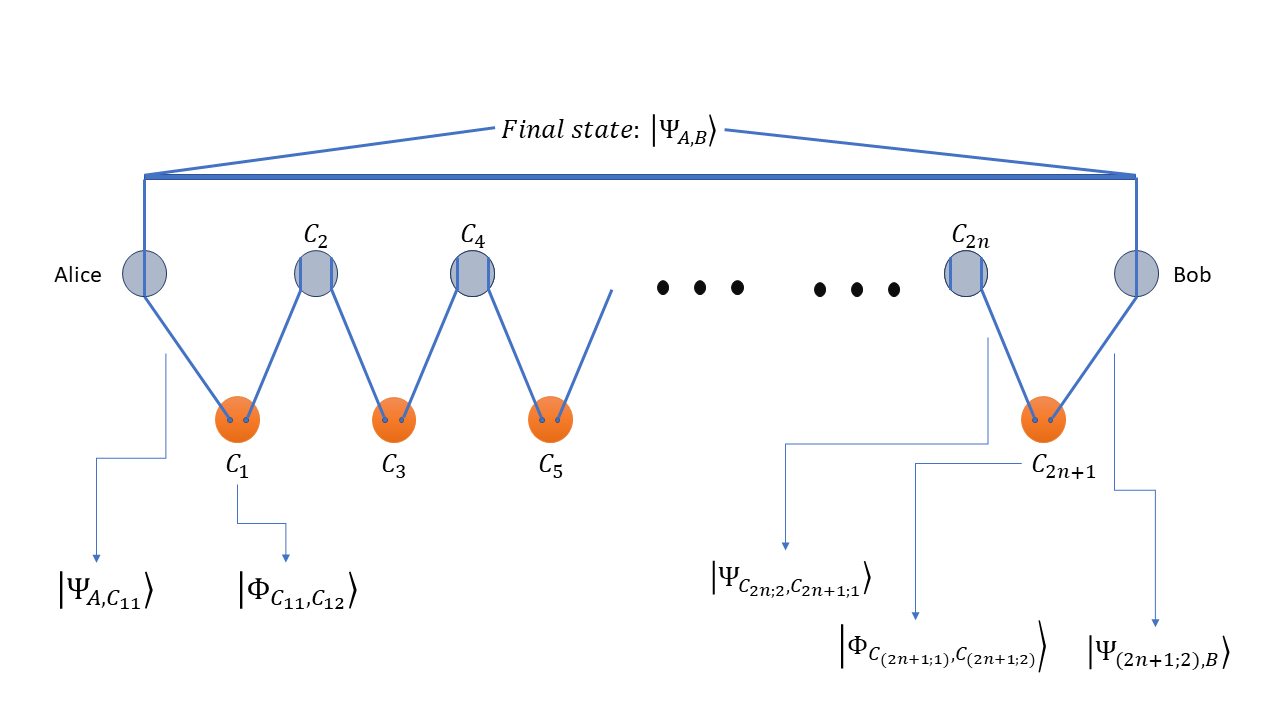}
    \caption{(Color online) This figure represents a quantum network with counterfactual quantum repeaters that can be implemented using counterfactual CNOT gates. $C_1,C_2, C_3, \cdots, C_{2n+1}$ are the nodes that are present between the extreme parties Alice and Bob. Alice's and Bob's photons are represented by the circle with a single line as they have one photon each. Now, the nodes $C_1, C_3, C_4,\cdots, C_{2n+1}$ possess entangled pairs of electrons each, and their respective entangled quantum states are represented by $\ket{\Phi_{C_{2i+1,1},C_{2i+1,2}}}$. For example, $\ket{\Phi_{C_{3,1},C_{3,2}}}$ represents the state where the third node's two electrons are entangled. Furthermore, $C_2, C_4, C_6,\cdots,C_{2n}$ are the nodes that possess a pair of uncorrelated photons each. These nodes are represented by circles with two straight lines, each line representing a photon. The joint state of the photon and the electron is represented by $\ket{\Psi}$. Here, the indices represent the photon and electron in the joint state. For example,  $\ket{\Psi_{C_{1,2},C_{2,1}}}$ represents the joint state of the first node's second electron and the second node's first photon.}
    \label{CQC_Fig2.PNG}
\end{figure*}

We consider a quantum channel between Alice and Bob who possess one photon each. Assuming there are $2n+1$ nodes within the channel, say $C_1, C_2, ... ,C_{2n+1}$, we consider that the nodes $C_1, C_3, C_5,...,C_{2n+1}$ consist of a pair of entangled electrons whereas the nodes $C_{2}, C_4, C_6,...,C_{2n}$ consist of a pair of photons. If the entanglement is distributed counterfactually among the nodes using a counterfactual CNOT gate such that one electron of $C_{2i-1}, i \in [0,n]$ is entangled with a photon $C_{2i}$ in the entire channel, then a repeater scheme can be visualized by making Bell measurement on the pair of electrons and then the pair of photons to establish entanglement between the photons with Alice and Bob. The scheme is illustrated clearly in Fig. \ref{CQC_Fig2.PNG}. This figure depicts these joint states between the electrons and photons as per our counterfactual scheme. Entanglement swapping via a perfect Bell measurement on the odd-numbered nodes takes place through the entangled electrons, and then linear optical Bell measurements (LOBMs) are employed to swap entanglement between multiple photons. LOBMs happen on the even-numbered nodes. So when a Bell measurement is made on $C_1$, Alice and the node $C_2$ get entangled counterfactually. The same thing happens with the successive nodes. Finally, linear optical Bell measurements are used to obtain the final entangled state with Alice and Bob. This is relatively better than repeaters with a setup that solely requires linear optical Bell measurements, as the probability of success in the latter case is extremely low. Moreover, the advantage of counterfactuality plays an important role in the entanglement distribution. It is to be noted that a perfect Bell measurement is indeed possible when dealing with a pair of electrons. However, if we restrict ourselves to linear optical Bell measurements, then the probability of success of Bell measurement on a pair of photons becomes $50\%$. While calculating the efficiency of the entire process, it is also important to consider that the counterfactual-CNOT gate's success probability is not exactly unity. We discuss this in the following subsection. 

\subsection{Performance analysis}
\subsubsection{Calculating the success probability}
The total number of nodes in the protocol is assumed to be $2n+1$. Alice and Bob are the parties that exist on the extreme ends of a channel with these $2n+1$ nodes. While distributing the entanglement between the entangled electrons (the odd-numbered node), one photon with its successive (even numbered) node, and another photon with its preceding node (even numbered), a counterfactual CNOT gate is used in a fashion that we have described earlier. Hence, every odd-numbered node makes use of two counterfactual-CNOT gates. This gives the total number of counterfactual-CNOT gates in the channel to be $2n+2$. Assuming that the probability of success of the $j^{th}$ counterfactual-CNOT gate is $P_j$, the probability of success of the $2n+2$ counterfactual-CNOT gates in the channel becomes a product of these success probabilities of each of these gates. It is to be noted that a total of $n$ linear optical Bell measurements will be required after the counterfactual entanglement swapping process. Incorporating the fact that linear optical Bell measurements can have a maximum success probability of $\frac{1}{2^n}$, the effective success probability becomes:
\begin{equation}
    P_{eff}=\frac{1}{2^n}\prod_{j=1}^{2n+2}P_j.
\end{equation}
As observed above, the increase in number of nodes will lead to a decrease in efficiency. However, this success probability can be steadily increased. In what follows, we will address this issue.
\subsubsection{Entanglement distribution time}
The time which is required to distribute the entanglement between adjacent electrons in the odd-numbered nodes, as well as the entanglement between the electrons and photons via the counterfactual-CNOT gates would be inversely proportional to the success probabilities. The generic average time for distribution of an entangled pair with $2n+1$ nodes is given by:\cite{JMPRA2012}

\begin{equation}
    T_{tot}=\Big(\frac{3}{2}\Big)^{2n+1} \frac{L_0}{c}\frac{1}{P_1 P_2 \cdots P_{2n+1}},
\end{equation}
where, $P_1, P_3, ..., P_{2n+1}$ are probabilities of successful entanglement swapping at the nodes with a pair of entangled electrons, whereas $P_2, P_4, \cdots, P_{2n}$ correspond to the probability of successful linear optical Bell measurement for entanglement swapping. $L_0$ refers to the length of the elementary link, i.e., the distance between two successive nodes. Depending on the success rate, the time required to distribute an entangled pair between the two parties can be derived. The factors crucial for an efficient system of entanglement swapping are the capabilities of detection, memory retention, and transmission. Represented by $\eta_D, \eta_M, \eta_t$ respectively\cite{JMPRA2012}, these efficiencies contribute to the total entanglement distribution time, which can be derived easily as:
\begin{equation}
T_{tot}=\frac{3^{2n+1}}{2^{n+1}}\frac{L_0}{c}\frac{1}{(\eta_D\eta_M)^{3n+2}\eta_t^{n+1}}
\label{CQC_eqn17}.
\end{equation}

\subsection{Possible improvements}

The efficiency is found to decrease drastically due to the limitations of linear optical Bell measurements. It can be increased if nonlinear optical methods are employed to measure the entangled quantum states. Another technique that may be used to enhance the efficiency is to use counterfactual Bell state analysis which is capable of analyzing the four Bell states, and thus, the only factor that would contribute to efficiency is the success probability of counterfactual CNOT gates.\cite{FZSR2018} Qubit amplifiers can also be used to obtain high entanglement distribution rates\cite{JMPRA2012}. 

\section{conclusion}
In this paper, we have shown that a counterfactual CNOT gate can be used to generate entanglement between an electron and a photon. Entanglement generated in such a manner is then used to develop a protocol for counterfactual entanglement swapping. Subsequently, the application of the protocol is analyzed in the regime of quantum repeaters with a specific focus on different aspects of success probabilities and entanglement pair distribution time. This paper intends to help to understand the applications of the counterfactual quantum theory in the domain of quantum communication. The protocol that we have proposed can not only be applicable for quantum repeaters but also in certain schemes of quantum cryptography for which the distribution of entangled states is necessary. We conclude the paper with the hope that this work will lead to the reporting of many such applications and their experimental realizations in the future.
\label{CQC_Sec5}

\section*{Acknowledgments}
AP acknowledges the support from the QUEST scheme of the Interdisciplinary Cyber-Physical Systems (ICPS) program of the Department of Science and Technology (DST), India, Grant No.: DST/ICPS/QuST/Theme-1/2019/6 (Q46). KT acknowledges support by the project OP JAC CZ.02.01.01/00/22{\_}008/0004596 of the Ministry of Education, Youth, and Sports of the Czech Republic. Authors also thank Dr. R. Srikanth for his useful inputs related to this work.

\end{document}